\documentclass[aps,prl,twocolumn,groupedaddress]{revtex4-1}
\usepackage{lipsum}
\usepackage{siunitx}
\usepackage{graphics}
\usepackage{epsfig}
\usepackage{color}
\begin{document}



\title{Generation of high energy laser-driven electron and proton sources with the  200 TW system VEGA 2 at the Centro de Laseres Pulsados}

\author{L. Volpe$^{1,2}$} 
\author{R. Fedosejevs$^{3}$}
\author{G. Gatti$^{1}$}
\author{J.A. P\'erez-Hern\'andez$^{1}$}
\author{C. M\'endez$^{1}$}
\author{J. Api\~naniz$^{1}$}
\author{X. Vaisseau$^{1}$}
\author{C. Salgado$^{1}$}
\author{M. Huault$^{1}$}
\author{S.  Malko$^{1}$}
\author{G. Zeraouli$^{1}$}
\author{V. Ospina$^{1}$}
\author{A. Longman$^{3}$} 
\author{D. De Luis$^{1}$}
\author{K. Li$^{1}$}
\author{O. Varela$^{1}$} 
\author{E. Garc\'ia $^{1}$} 
\author{I. Hern\'andez$^{1}$} 
\author{J.D. Pisonero$^{1}$} 
\author{J. Garc\'ia Ajates$^{1}$}
\author{J. M. Alvarez$^{1}$}
\author{C. Garc\'ia$^{1}$}
\author{M. Rico$^{1}$}
\author{D. Arana$^{1}$}
\author{J. Hern\'andez-Toro$^{1}$}
\author{L. Roso$^{1,2}$}

\affiliation{$^{1}$Centro de L\'aseres Pulsados (CLPU), Edificio M5. Parque Cient\'ifico. C/ Adaja, 8. 37185 Villamayor, Salamanca, Spain}
\affiliation{$^{2}$Universidad de Salamanca, Salamanca, Spain}
\affiliation{$^{3}$Department of Electrical and Computer Engineering,University of Alberta, Edmonton Alberta, Canada T6G 2V4}

\begin{abstract}
The Centro de Laseres Pulsados in Salamanca Spain has recently started operation phase and the first User access period on the 6 J 30 fs 200 TW system (VEGA 2) already started at the beginning of 2018. In this paper we report on two commissioning experiments recently performed on the VEGA 2  system in preparation for the user campaign. VEGA 2  system has been tested in different configurations depending on the focusing optics and targets used. One configuration (long focal length f=130 cm) is for under-dense laser-matter interaction where VEGA 2  is focused onto a low density gas-jet generating electron beams (via laser wake field acceleration mechanism) with maximum energy up to 500 MeV and an X-ray betatron source with a 10 keV critical energy. A second configuration (short focal length f=40 cm) is for over-dense laser-matter interaction where VEGA 2 is focused onto an 5 $\mu$m thick Al target generating a proton beam with a maximum energy of 10 MeV and average energy of 7-8 MeV and temperature of 2.5 MeV.  In this paper we present preliminary experimental results.
\end{abstract}

\keywords{High Power Laser; laser-plasma; particle acceleration}

\maketitle

\setcounter{section}{0} 
\section*{Introduction}

Laser technology has advanced to the point where hitherto unobtainable intensities are now routinely achievable, and rapid progress is being made to increase intensities further \cite{mourou1985}. The versatility of high power lasers has resulted in their use in a broad range of scientific fields, including novel particle accelerators, fusion research, laboratory astrophysics, condensed matter under high pressure, novel x-ray sources, and strong-field QED, amongst others. The potential for developing compact, high brightness particle and radiation sources has given a strong impetus to the development of the underpinning laser technology, including increasing the efficiency and repetition rate of the lasers. A result of this technological development can be seen in the new generation of ultrafast high power laser systems working at High Repetition Rate (HRR) which have been built across Europe \cite{collin2015}. One of the more relevant and representative is the Centro de Laseres Pulsados \cite{clpu} in Salamanca Spain which has recently started operation. The CLPU has been founded by Spanish Ministry  of Economy, Junta de Castilla y Le\'on and the University of Salamanca and its main system  VEGA consist in a 30 fs pulse delivered in three different arms of 20 TW (VEGA 1), 200 TW (VEGA 2) and 1 PW (VEGA 3).  CLPU has recently started operation phase, the first User access period on the VEGA 2 already started at the beginning of 2018 and a commissioning experiment on VEGA 3 is planned for 2019.  VEGA 2 has been previosly tested in different configurations depending on the focusing optics and targets used. One configuration is designed for under-dense laser-matter interaction where VEGA 2 is focused (F=130 cm , $\phi_L$=20 $\mu$m, $Z_r$= 260 $\mu$m ) onto a low density gas-jet generating (via wake field mechanism \cite{tajima1979,mangles2004,faure2004,geddes2004,malka2012}) electron beams with maximum energy up to 500 MeV and an X-ray betatron source with 10 keV characteristic critical energy \cite{albert2008,kneip2010,mo2013,albert2016}. The second configuration is designed for overcritical density laser-matter interaction where VEGA 2 is focused ((F=40 cm , $\phi_L$=7 $\mu$m, $Z_r$= 25 $\mu$m) ) onto a 5 $\mu$m Al target generating (via TNSA mechanism \cite{wilks2001,maksimchuk2000,clark2000,snavely2000}) a proton beam with a maximum energy of 9 MeV and average temperature of 2.5 MeV. In this paper both the commissioning experiments are reported and explained by describing the experimental set-up and showing the capabilities of the VEGA system. Finally preliminary results are also given. 

\section{The VEGA system} 
\begin{figure}[h]
\centering
\includegraphics[angle=0,scale=0.5]{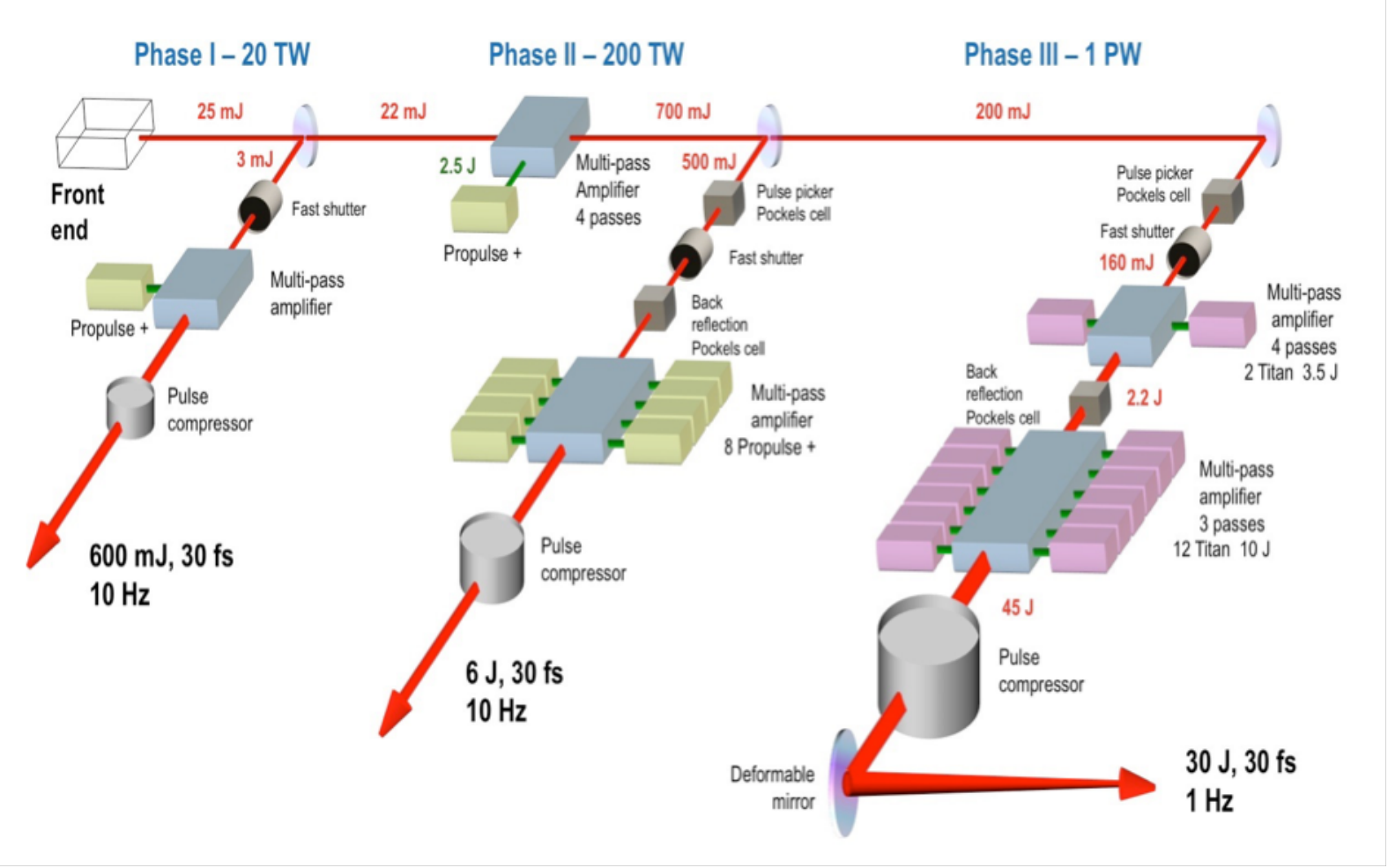}
\caption{Scheme of the three VEGA systems}
\label{fig_1}
\end{figure}

\begin{figure}[h]
\centering
\includegraphics[angle=0,scale=0.9]{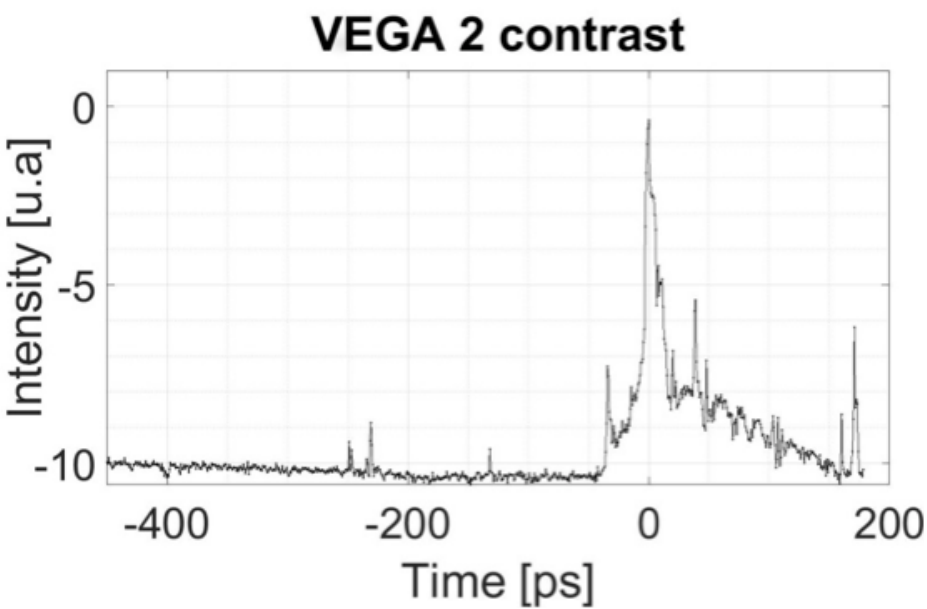}
\caption{Contrast of VEGA 2, measured by Sequioa.}
\label{contrast_laser}
\end{figure}

The VEGA laser is a CPA Titanium:sapphire system, with a central wavelength of 800 nm $\pm$ 10 nm. It has three arms (see fig.\ref{fig_1}) with maximum power 20 TW (VEGA-1), 200 TW (VEGA-2) and 1 PW (VEGA-3). 
The common front-end has a double CPA as well as an XPW system that increases the contrast of the pulses significantly, as shown in Fig.\ref{contrast_laser} , making  it suitable for high density targets. 

The temporal contrast is: @ ns (prepulse) 5 $\times 10^{-10}$;  @ 1 ps 2 $ \times 10^{-5}$;  @ 5 ps 5 $\times 10^{-8}$;  @ 10 ps 8  $\times 10^{-9}$; @ 100 ps 5 $\times 10^{-12}$. Fig.$\ref{contrast_laser}$ shows a measuremnet of VEGA laser contrast just after the compressor. 

In April  2017 the first call for USER access on the VEGA 2 system was lunched by CLPU, thirty Scientific proposals were received for a total amount of more than 500 days requested over the 100 days offered in the call. The VEGA 2 system has been offered at high repetition rate with some limitations mainly connected to targetry and diagnostics. In preparation of such user access we organised a series of commissioning experimental campaigns along 2017. The main goal of this campaign was to learn how to prepare the VEGA 2 system for future user access campaigns. In particular we prepared the VEGA 2 target area to be easily adaptable for different possible experimental setups by changing the laser focusing system and by implementing a simple system to split the main beam (via beam splitters) at different fractions among which  1/99 \% and 10/90 \%;  other possibilities such as  50/50 \%  are under development. Depending on the experimental requirements a probe beam can be focused by one of the parabolic mirrors or simply by optical lenses. Both the parabolic mirrors are protected by thin fused silica pellicles placed just in front of them. Laser pulse duration is routinely measured close to the target position by using a second order auto-correlator and compared with second harmonic optimisation measurements. 


\section{The Commissioning experimental campaign}
The VEGA2 target area currently offers two different configurations with the possibility of several laser probes depending on the experimental requests. For both the configurations the laser VEGA 2 has been delivered at 30 fs, with a maximum power of 150 TW corresponding to a maximum energy at the entrance of the compressor of around 7 Joules. The conversion efficiency $\eta$ from the entrance of the compressor up to the target is measured to be around 50\% giving a total energy on target around 3.5 J.  Assuming the same energy conversion efficiency for both the configurations the total laser intensity  is estimated to be $I_L[W/cm^2] \sim 5.8 \times 10^{21}/A[\mu m^2] $, where  $A[\mu m^2]= \pi r_L^2$.

\subsection{Long focalisation in gaseous targets} 

\begin{figure}[h]
\centering
\includegraphics[angle=0,scale=0.5]{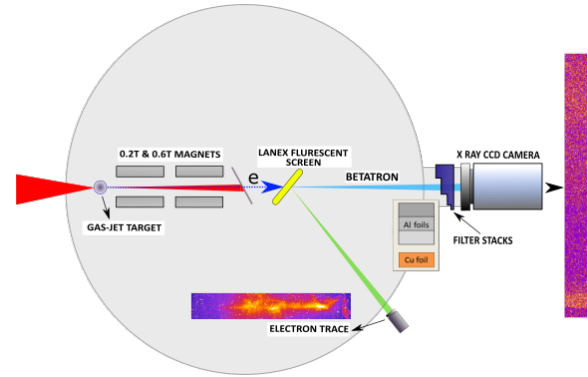}
\caption{Long focal experimental setup}
\label{fig_lf}
\end{figure}

\begin{figure}[h]
\centering
\includegraphics[angle=0,scale=0.30]{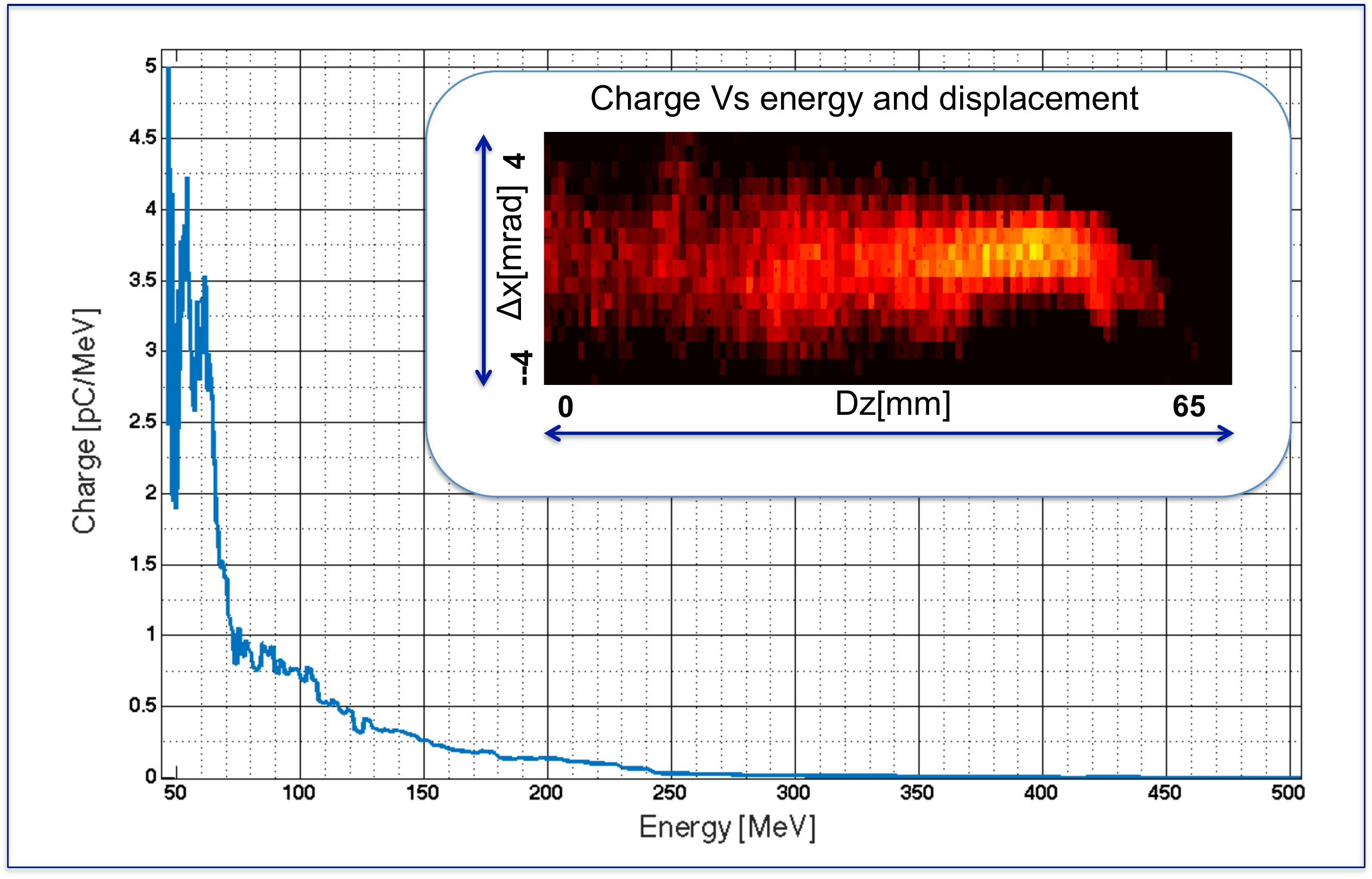}
\caption{(main image) typical non-monoenergetic electron spectrum measured with 1.2 tesla magnetic spectrometer; (insertion) filtered image of measured electron energy spectrum}
\label{e_spec}
\end{figure}

VEGA 2 was focused by an F/13 parabolic mirror into a 5 mm thick gas-jet placed at 30 cm far from the TCC, as shown in fig. $\ref{fig_lf}$. The  laser beam waist has been measured (at low power) to be  FWHM 17.7 $\mu$m  $\pm$  0.5  $\mu$ m with less then $ 50 \%$ of the total energy in the FWHM giving a peak intensity within the FWHM of around $ 1.1 \times10^{19}$ W/cm$^2$. Pulse duration has been adjusted during the experimental campaign to find the best acceleration performance by considering the relation between plasma frequency and laser pulse duration.  Fully ionized plasma density has been fixed to $1.2 \times 10^{19}$ cm$^{-3}$ (over one order of magnitude larger than the threshold for the bubble regime)  to maximise betatron radiation emission by generating a broad electron beam energy spectrum and leading to an increased accelerated charges (see fig. $\ref{e_spec}$).

Electron beams up to 500 MeV have been measured with an electron spectrometer composed by different magnet dipoles coupled with a Lanex scintillating detector and an imaging system (see fig. \ref{fig_lf}). Betatron radiation has also been measured with  typical Synchrotron-like energy spectrum characterised by a critical energy of $\approx$ 10 keV. Betatron spectrum have been analysed with an X-ray CCD camera by using Ross filters technique. We estimated to have a peak brightness greater than $10^{8}$ photons/srad/0.1$\%$BW/shot on average, with a divergence greater than $9.2 \times 4.5$ mrad in horizontal and vertical directions. This Betatron source has been also used as high frequency (short wavelength) probe to investigate a pre heated  Warm Dense Aluminium sample and experimental results are now under analysis and will be published soon.


 
\subsection{Short focalisation in solid targets} 
VEGA 2 was focused by an F/4 gold coated parabolic mirror onto 5 $\mu$m thick Al foil with an angle of 10 degrees with respect to the normal of the foil. 
The  laser beam waist has been measured (at low power) to be  FWHM 8 $\mu$m  $\pm$  2  $\mu$m with 50 \% of the total energy in the FWHM giving a peak laser intensity within the FWHM of  around $I_L \simeq 2 \times 10^{20} \;\text{W/cm}^2$. Laser pulse duration  has been routinely  measured to be 30 fs $\pm$ 3 fs. A three axis motorised target holder has been used holding 5 and 10 $\mu$m thick Al foils. Target alignment was done prior to each shot with a resolution of around 5 $\mu$m  ($\le Z_r ~ 25 \mu$m)

\begin{figure}[h]
\centering
\includegraphics[angle=0,scale=0.5]{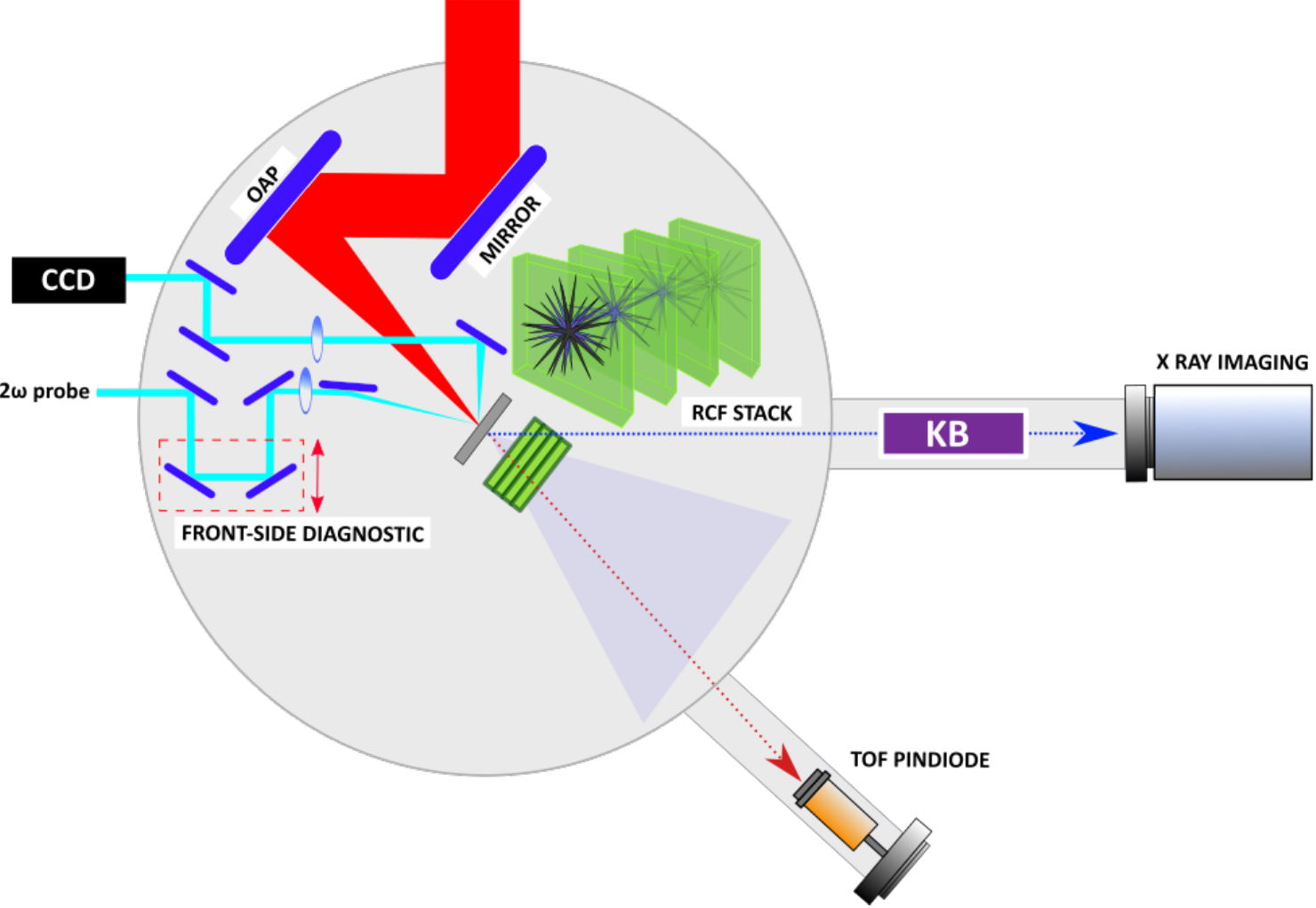}
\caption{Short focal experimental setup }
\label{fig_sf}
\end{figure}

\begin{figure}[h]
\centering
\includegraphics[angle=0,scale=0.53]{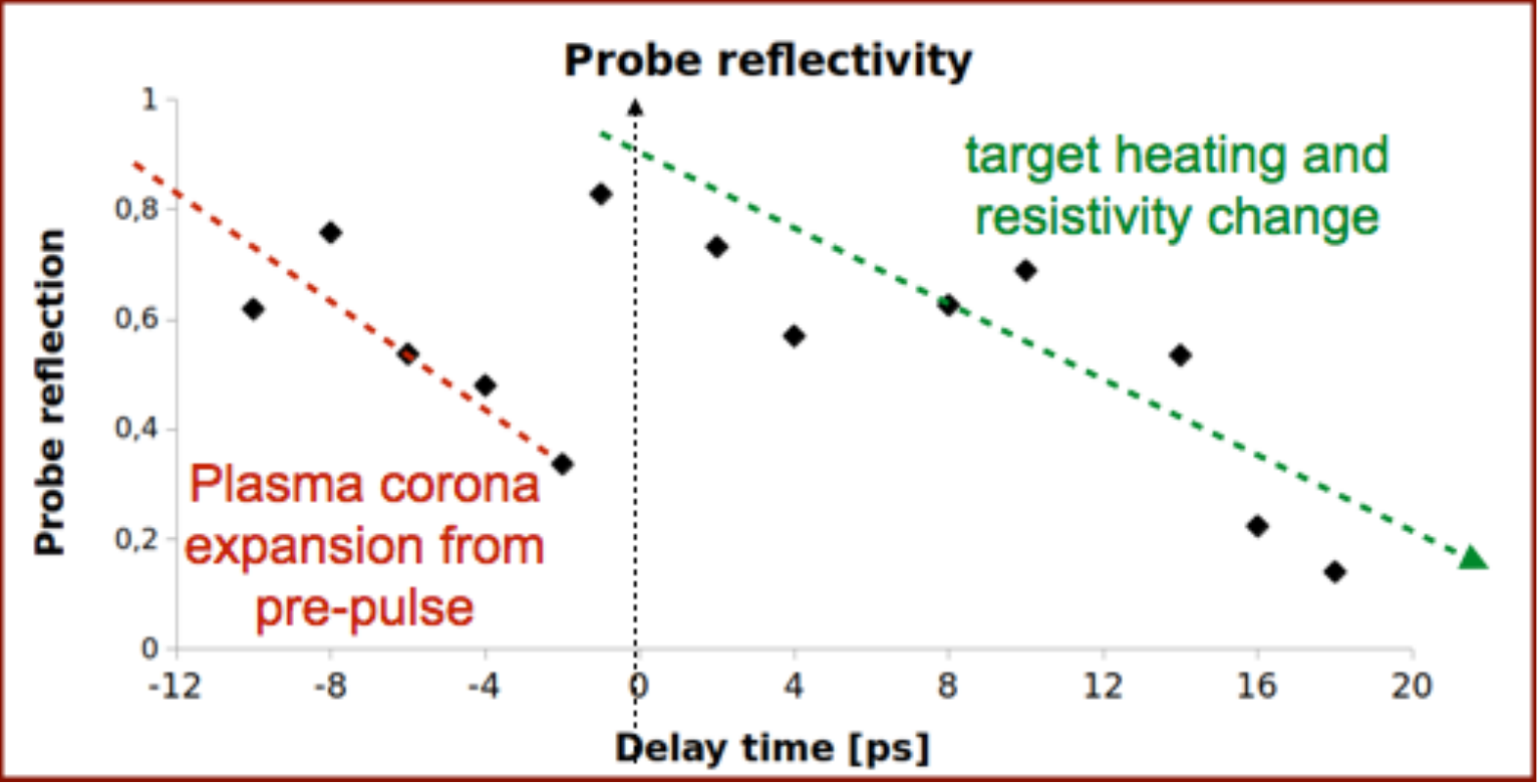}
\caption{Normalized reflection of the probe at the target for different time delays with VEGA 2. doted lines are guides for eyes}
\label{contrast}
\end{figure}

For such laser intensities and so thin targets the control of laser prepulse \cite{kaluza2004}  and plasma corona expansion is mandatory. A first measurement of the laser contrast just after the compressor has been done by laser team and  it is shown in fig. $\ref{contrast_laser}$. We developed a simple front-side diagnostic to control the laser-target interaction by looking at plasma formation as a function of time in a time range of $\pm$ 300 ps with respect the main beam arrival. A small part of the main beam is extracted, doubled in frequency, its polarization rotated $\ang{90}$ with respect the main laser and focused (couterpropagating with respect to the main laser pulse) on the front target surface with an angle of $\approx \ang{20}$. The reflected beam was imaged onto an optical CCD camera, while a part of the transmitted beam is also recorded for reference. The setup is shown in fig.$\ref{fig_sf}$.
Comparison of the amount of reflected probe beam at different delays will help for a preliminary interpretation and understanding of the laser contrast efficiency in laser-solid interaction; a further optimisation of the diagnostic method will help to reconstruct the plasma density profile before during and after the main pulse arrival by comparison to simulations. Fig. $\ref{contrast}$ shows reflected fraction of the probe as a function of the chosen delay.  

\begin{figure}[h]
\centering
\includegraphics[angle=0,scale=0.35]{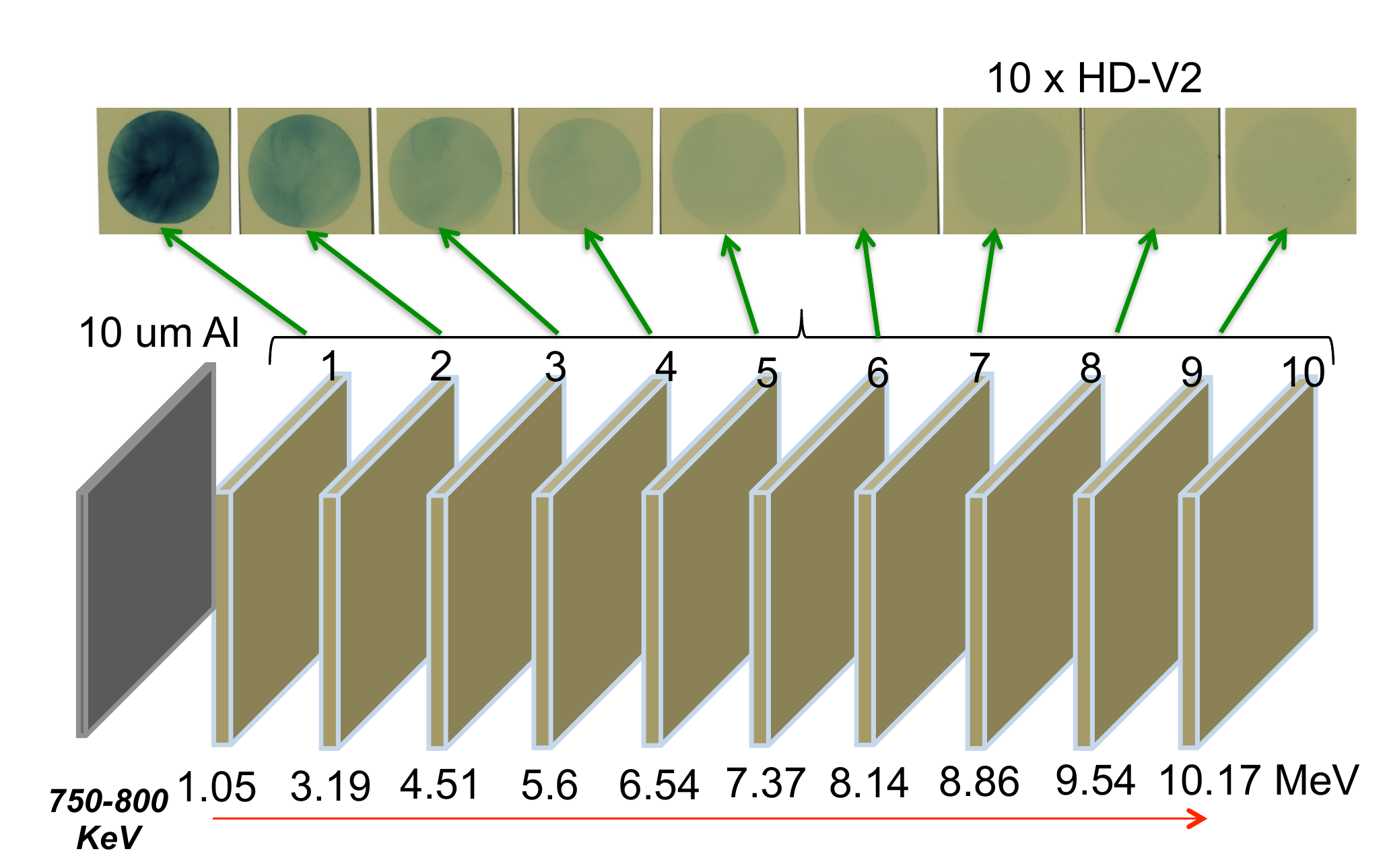}
\caption{RCF stack design and experimental results}
\label{fig_rcf}
\end{figure}

\begin{figure}[h]
\centering
\includegraphics[angle=0,scale=0.38]{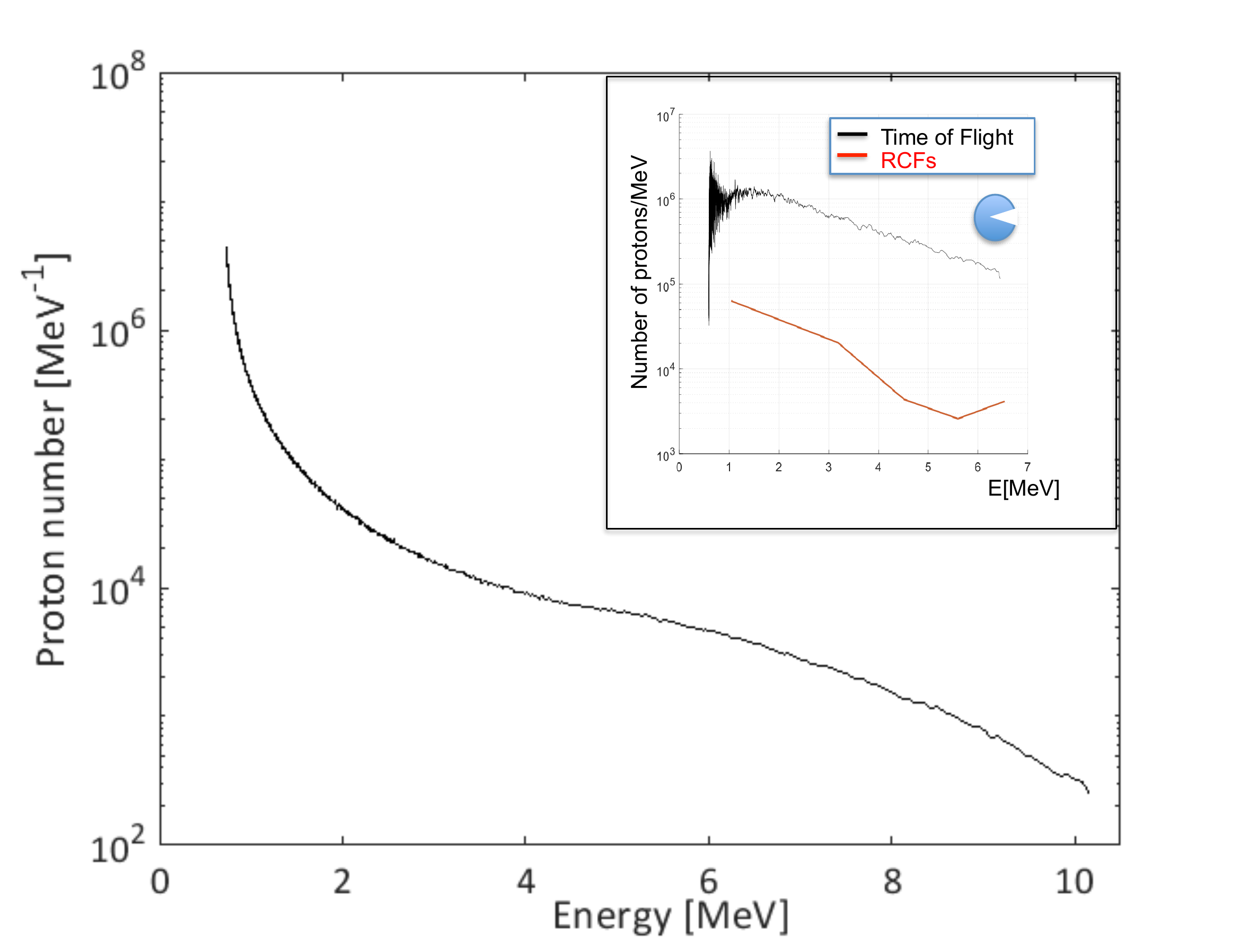}
\caption{Proton spectrum obtained form typical Time of Flight measurements ; In the insertion a proton spectrum measurement with partially cut RCFs of  (black line) ToF and (red lines) RCF stack }
\label{proton-spectrum}
\end{figure}

Proton beam energies up to 10 MeV have been routinely measured with different diagnostic techniques. Proton spatial and energy distribution have been measured firstly with a stack of Radio Chromic Films (RCFs)  with the energy resolution limited by the number and thickness of RCF layers to $\Delta E \sim 0.5 MeV$.  
Time of Flight (ToF) measurements have been performed by using 1 ns  time resolved pin diode detector  and with a 300 ps time-resolved Micro Channel Plate (MCP) both placed 2.5 Meter far from the Target Chamber Center (TCC) as shown in Fig. \ref{fig_sf}. Proton measurements were also checked by CR39  detectors placed beside the RCF stack or the ToF detectors. The measurements show TNSA-like proton energy spectrum with an averaged (the average is done over different acquisitions) mean temperature $\langle T_{mean} \rangle  \sim 2.5 $  MeV, an averaged maximum temperature $\langle T_{max} \rangle  \sim 9$ MeV and a divergence angle ranging between $\ang{20}- \ang{25}$ FWHM. Fig. \ref{fig_rcf} shows the RCF stack scheme with an example of experimental results. Fig. \ref{proton-spectrum} shows proton energy spectrums obtained by analysing both  RCFs stack  and Time of Flight detectors. The Spectrum obtained by RCFs analysis \ref{volpe2011,valerianew} .. give larger number of protons compared to the on e obtained from ToF analysis. This difference relies on the fact that the two detectors have different sensitivity.  

\begin{figure}[h]
\centering
\includegraphics[angle=0,scale=0.40]{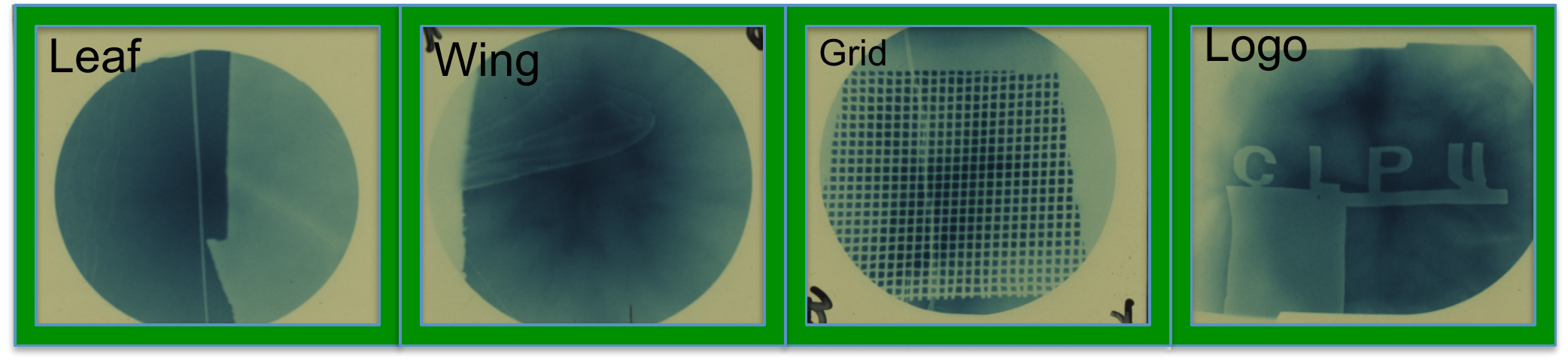}
\caption{Proton radiography of (starting from left) a peace of leaf, a wing, a metallic grid and a CLPU metallic logo. The final image of the logo appear in the first RCF layer to be in  a 20 mm x 100 mm area which, according to the geometrical magnification M $\sim$ 4  reproduce correctly the original  dimensions of the logo which was 5 mm x 20 mm }
\label{fig_rad}
\end{figure}

Proton radiography of test objects has been performed with the obtained proton distribution and some results are shown in fig. $\ref{fig_rad}$ starting from left a pieces of a leaf, a wing, a grid and the CLPU metal logo being radiographed.  Analysis of the results are in progress and will be soon submitted for publication. 

TNSA proton acceleration mechanism is controlled by electron dynamics into the target and it is relevant to also have information on the laser-driven electron population \cite{volpe2013} which generates the quasi-static electric field \cite{schreiber2006,passoni2008} responsible for the proton acceleration. We imaged the electron beam travelling into the target by collecting and focusing incident K alpha radiation coming from hot electron interaction within the Al target ($h \omega \sim $1.5 KeV). Such radiation has been routinely collected by a remotely adjustable Kirk-Patrick Baez (KB) microscope, onto an X-ray CCD. The KB microscope was placed 1.2 m from TCC at $\ang{30}$  with respect to the target normal and the CCD camera placed in the KBs focal plane. This compact X-ray K alpha diagnostic has also been tested by performing radiography of a calibration grid. Fig. 8 shows the diagnostic set up (a) an example of an obtained K alpha spot (b) and Results from X ray radiography of a grid pattern (c).

\begin{figure}[h]
\centering
\includegraphics[angle=0,scale=0.30]{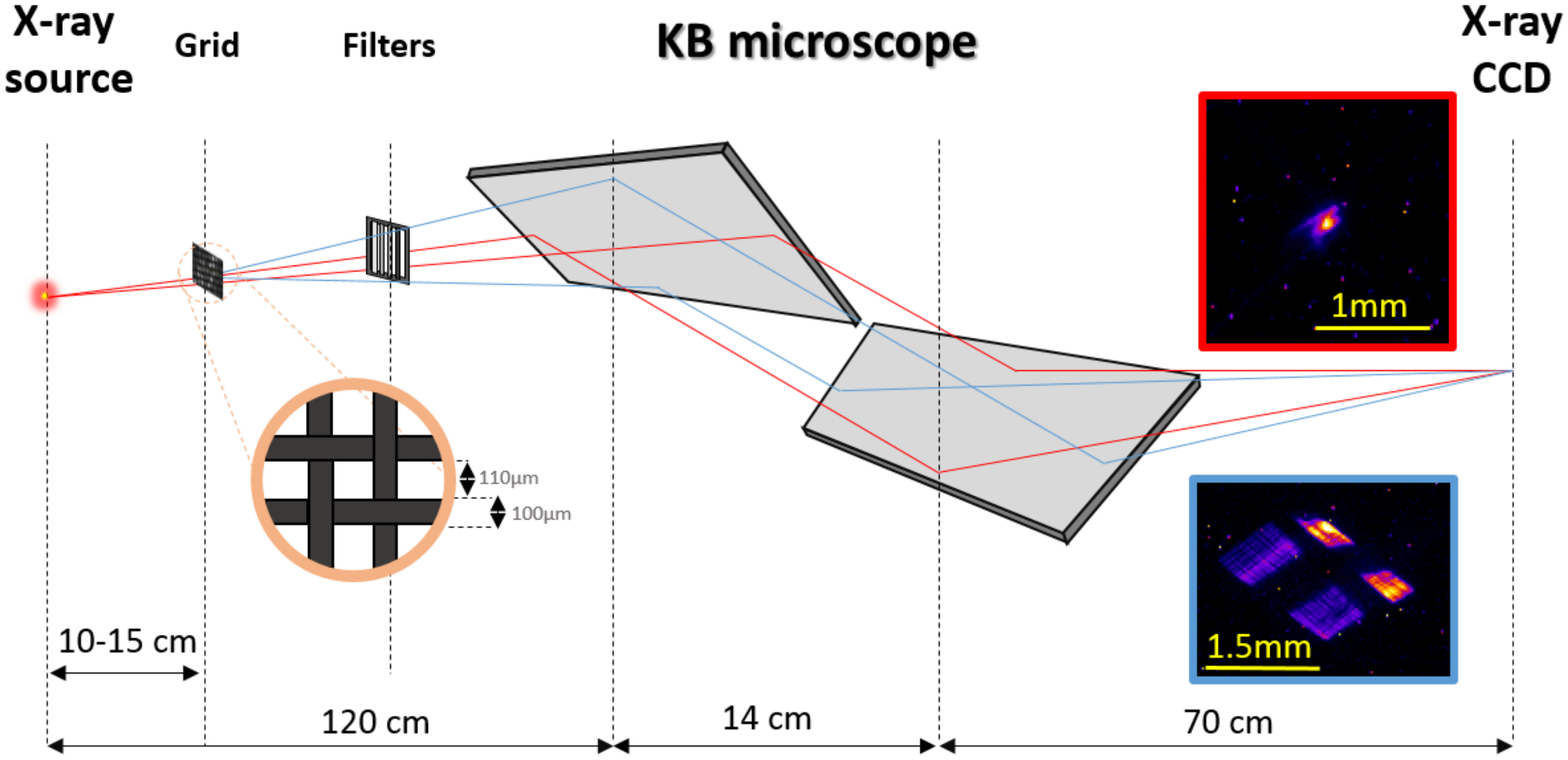}
\caption{(main image) KB microscope set-up; (insert top right) image of the k alpha emission from fast electron beam travelling into the 6 um Al target. (insert bottom right) Magnified radiography of a calibration grid}
\label{fig_kb}
\end{figure} 

\begin{figure}[h]
\centering
\includegraphics[angle=0,scale=0.55]{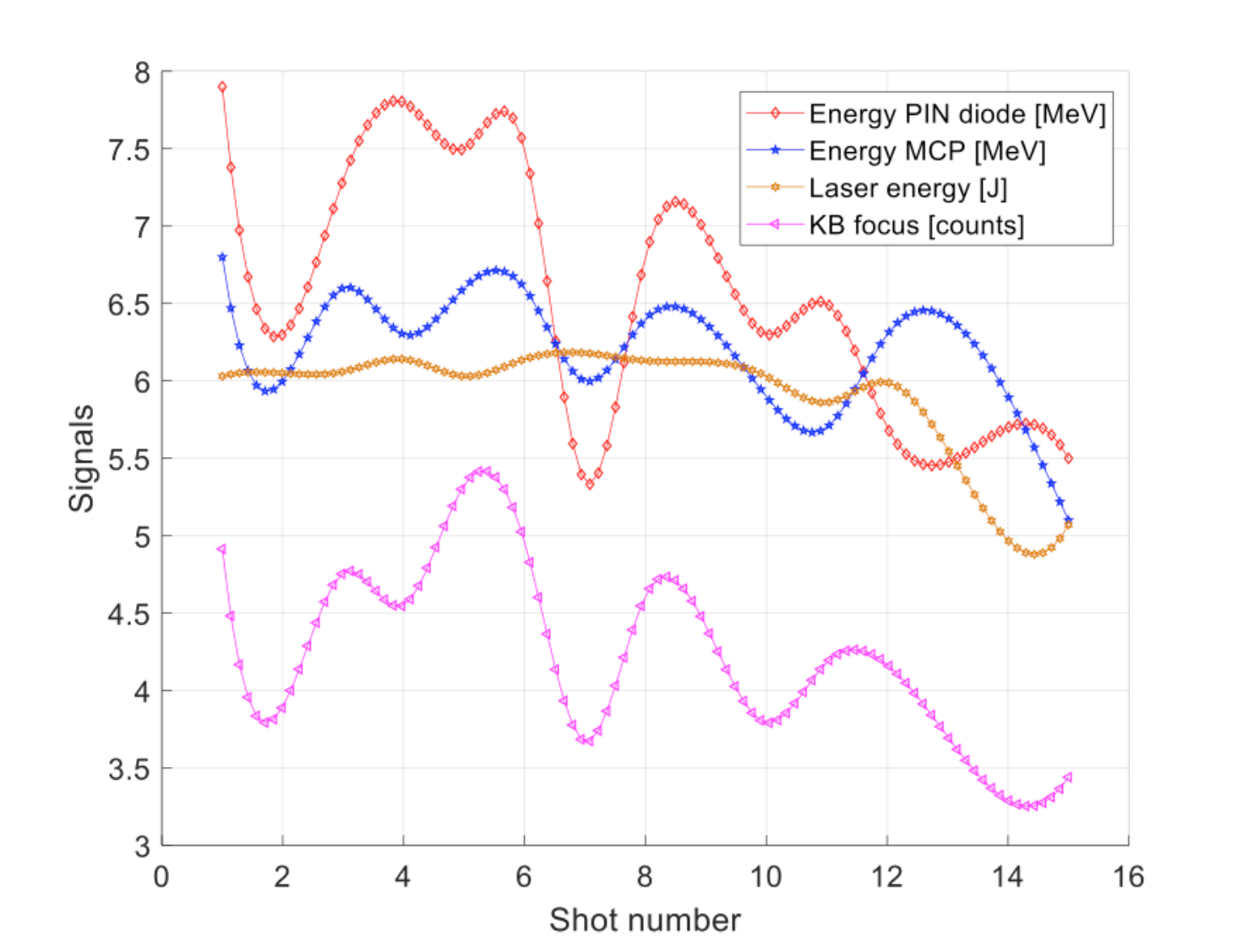}
\caption{Example of experimental data for different diagnostics in place:  (red, diamond points) maximum proton energy from pin diode ToF measurement, (blue, stars points ) maximum proton energy from MCP TOF measurements, (orange, circle points) Laser energy and (purple, triangle points) K-alpha integrated signal from KB Xray focusing system. The lines are guide for eyes}
\label{fig_ave}
\end{figure}

Proton production has been tested at a constant repetition rate mode and the VEGA 2 laser pulse operation has shown good stability in terms of shot to shot variation of the focal spot and total laser energy which results in a good reproducibility of the experimental data in all the diagnostics. The K alpha signal was used as a reference criteria for evaluating shot-to-shot fluctuation. Fig.\ref{fig_ave} shows an example of a series of shots comparing experimental data acquired from different installed diagnostics Pin diode, Multi-Channel-Plates and KB microscope. Clear correlation between results can be seen.


\section{Conclusions}
The first experimental campaigns on VEGA 2 target area have shown the potential of the 200 TW system demonstrating a good performance and stability of the VEGA 2 laser system with a large  margin of improvement. A Flexible  and comprehensive  targetry and diagnostic system have been developed to support experimental activities in both long focal and short focal lengths configurations. 

It is worth noting  that  targetry and diagnostic systems need further development to match the repetition rate potential of the laser system. Indeed  target movement and alignment still need to be improved and optimised for working modes above 0.1 Hz. Also diagnostic techniques are not always mature for high repetition rate acquisition that can reproduce the performance of the original passive detectors.  
In summary we have routinely obtained 0.5 GeV electron beams with more then 10 KeV critical energy (synchrotron-like spectrum) X-ray betatron emission in the long focal length configuration and a TNSA-like Proton spectrum with maximum energy of around 9 MeV in the short focal length configuration. In addition Significant Electro Magnetic pulses have been also generated and  measured during laser operation, in particular during the short focal campaign where laser intensities reach up to $10^{20}$ W/cm$^2$. 

\vspace{6pt} 



\section*{Acknowledgments}
The authors would like to thanks all the CLPU workers that during the last 10 years were working hard in order to built and develop the facility. In particular the CLPU scientific, laser and administrative divisions as well as the engineering and radio protection areas. Also we would like to thanks the Ministry of science, the Castilla y Leon region and the University of Salamanca for the constant support during all the phases of construction and implementation of the CLPU facility. 
Support from Spanish Ministerio de Ciencia, Innovación y Universidades through the PALMA Grant No. FIS2016-81056-R, ICTS Equipment Grant No. EQC2018-005230-P; from LaserLab Europe IV Grant No. 654148, and from Junta de Castilla y León Grant No. CLP087U16 is acknowledged.

\section*{Author contributions}
R. Fedosejevs and L. Volpe  conceived, designed and oversaw respectively the electron acceleration and the proton acceleration experiments; G. Gatti and J.A. Perez supervise and operated both the experiments;  X. Vaisseau, J. I. Api\~naniz, M. Huault, C. Salgado, G. Zeraouli, S. Malko, A. Longman, K. Li, V. Ospina,  performed the experiments and analysed the data; C. Mendez and Laser team operated the laser and D. de Luis  with Engineer area support experimental campaign with design and fabrication support. Luca Volpe wrote the first draft of the paper.

\appendix{}


\begin{thebibliography}{00}




\bibitem{mourou1985} D. Strickland and G. Mourou, Compression of ampli¯ed chirped optical pulses,
Optics Communications 56, 219 (1985).
\bibitem{collin2015}  C. Danson, D. Hillier1, N. Hopps, and D. Neely, High Power Laser Science and Engineering, (2015), Vol. 3, e3, 14 pages
\bibitem{clpu} www.clpu.es

 



\bibitem{tajima1979} T. Tajima and J.M. Dawson, Laser Electron Accelerator, Phys. Rev. Lett. 43, 267-270 (1979)
\bibitem{mangles2004} Mangles S P D et al Monoenergetic beams of relativistic electrons from intense laser-plasma interactions Nature 431 5358 (2004)
\bibitem{faure2004} Faure J, Glinec Y, Pukhov A, Kiselev S, Gordienko S, Lefebvre E, Rousseau J-P, Burgy F and Malka V 2004 A laser-plasma accelerator producing monoenergetic electron beams Nature 431 5414 (2004)
\bibitem{geddes2004} Geddes C G R, Toth Cs, Van Tilborg J, Esarey E, Schroeder C B, Bruhwiler D, Nieter C, Cary J and Leemans W P 2004 High-quality electron beams from a laser wakefield accelerator using plasma-channel guiding, Nature 431 53841 (2004)
\bibitem{malka2012}V. Malka, Laser Plasma Accelerators, Phys. of Plasmas 19, 055501 (2012)

\bibitem{albert2008}F. Albert et al., Betatron oscillations of electrons accelerated in laser wakefields characterized by spectral x-ray analysis, Phys. Rev. E 77,056402-1-(2008)
\bibitem{kneip2010}S. Kneip et al,, Bright spatially coherent synchrotron X-rays from a table-top source, Nature Physics 6, pages 980?983. (2010)
\bibitem{mo2013}M.Z. Mo et al., Laser wakefield generated X-ray probe for femtosecond time-resolved measurements of ionization states of warm dense aluminum Rev. of Scientific Instrum. 84, 123106-1-11 (2013)
\bibitem{albert2016} F. Albert, Applications of laser wakefield accelerator-based light sources, Plasma Physics and Controlled Fusion 58, 103001-1-35 (2016).




\bibitem{wilks2001} S.C. Wilks, A.B. Langdon, T.E. Cowan, M. Roth, M. Singh, S. Hatchett, M.H. Key, D. Pennington, A. MacKinnon and R.A. Snavely, Energetic proton generation in ultra-intense lasersolid interactions, Phys. Plasmas 8, 542 (2001).
\bibitem{maksimchuk2000} A. Maksimchuk, S. Gu, K. Flippo, D. Umstadter and V.Yu. Bychenkov, For- ward Ion Acceleration in Thin Films Driven by a High-Intensity Laser, Phys. Rev. Lett. 84, 4108 (2000).
\bibitem{clark2000}  E.L. Clark, K. Krushelnick, J.R. Davies, M. Zepf, M. Tatarakis, F.N. Beg, A. Machacek, P.A. Norreys, M.I.K. Santala, I. Watts and A.E. Dangor, Mea- surements of Energetic Proton Transport through Magnetized Plasma from Intense Laser Interactions with Solids, Phys. Rev. Lett. 84, 670 (2000).
\bibitem{snavely2000} R.A. Snavely, M.H. Key, S.P. Hatchett, T.E. Cowan, M. Roth, T.W. Phillips, M.A. Stoyer, E.A. Henry, T.C. Sangster, M.S. Singh, S.C. Wilks, A. MacK- innon, A. O®enberger, D.M. Pennington, K. Yasuike, A.B. Langdon, B.F. Lasinski, J. Johnson, M.D. Perry and E.M. Campbell, Intense High-Energy Proton Beams from Petawatt-Laser Irradiation of Solids, Phys. Rev. Lett. 85, 2945 (2000).




\bibitem{volpe2011} L. Volpe, D. Batani, B. Vauzour, Ph. Nicolai, J.J. Santos, C. Regan, A. Morace, F. Dorchies, C. Fourment, S. Hulin, Perez, S. Baton, K. Lancaster, M. Galimberti, R. Heathcote, M.Tolley, Ch.Spindloe, P. Koester, L. Labate, L.A. Gizzi, C. Benedetti, A. Sgattoni, M. Richetta, J. Pasley, F. Beg, S. Chawla, D.P. Higginson, A.G. MacPhee; "Proton radiography of laser-driven imploding target in cylindrical geometry" Phys. Plasmas 18, 006101 (2011)
\bibitem{volpe2013} L. Volpe, D. Batani, G. Birindelli,, A. Morace, P. Carpegiani, M. H. Xu, F. Liu,  Y. Zhang, Z. Zhang, X. X. Lin, F. Liu, S. J. Wang,  P. F. Zhu, L. M. Meng,  Z. H. Wang, Y. T. Li,  Z. M. Sheng, Z. Y. Wei, J. Zhang,  J.J. Santos, C. Spindloe; "Laser-driven Electron beam in matetr" Phys. Plasmas 20, 033105 (2013)
\bibitem{kaluza2004}  M. Kaluza, J. Schreiber, M.I.K. Santala, G.D. Tsakiris, K. Eidmann, J. Meyer ter Vehn and K.J. Witte, In°uence of the Laser Prepulse on Proton Acceleration in Thin-Foil Experiments, Phys. Rev. Lett. 93, 045003 (2004).
\bibitem{schreiber2006} J. Schreiber, F. Bell, F. GrÄuner, U. Schramm, M. Geissler, M. SchnÄurer, S. Ter-Avetisyan, B.M. Hegelich, J. Cobble, E. Brambrink, J. Fuchs, P. Aude- bert and D. Habs, An Analytical Model for Ion Acceleration by High-Intensity
Laser Pulses, Phys. Rev. Lett. 97, 045005 (2006).
\bibitem{Passoni2008} M. Passoni and M. Lontano Phys. Rev. Lett. 101, 115001 (2008)



 





\bibitem{mo2017}M.Z. Mo et al., Measurements of ionization states in warm dense aluminum with Betatron radiation,  Phys. Rev. E. 95, 053208 (2017)



\end{thebibliography}
\end{document}